\documentclass[pre,twocolumn,preprintnumbers,amsmath,amssymb,superscriptaddress,reprint]{revtex4} 

\usepackage{graphicx}
\usepackage{dcolumn}
\usepackage{bm,amsmath}
\usepackage{overpic}

\begin{document}
\title{One-dimensional hydrodynamic model generating a turbulent cascade}
\author{Takeshi \surname{Matsumoto}}
\email{takeshi@kyoryu.scphys.kyoto-u.ac.jp}
\affiliation{%
Division of Physics and Astronomy,
Graduate School of Science,
Kyoto University,
Kyoto, 606-8502, Japan
}
\author{Takashi \surname{Sakajo}}
\email{sakajo@math.kyoto-u.ac.jp}
\affiliation{%
Department of Mathematics,
Graduate School of Science,
Kyoto University,
Kyoto, 606-8502, Japan}
\date{\today}
\begin{abstract}
As a minimal mathematical model generating cascade analogous to that of the Navier-Stokes 
turbulence in the inertial range, 
we propose a one-dimensional partial-differential-equation model that 
conserves the integral of the squared vorticity analogue (enstrophy) in the inviscid case.
With a large-scale random forcing and small viscosity, we find numerically that the model exhibits the enstrophy cascade, 
the broad energy spectrum with a sizable correction to the dimensional-analysis prediction,
peculiar intermittency and self-similarity in the dynamical system structure.
\end{abstract} 


\maketitle

\section{Introduction}
One way to tackle the problem of fluid turbulence is to study
its model with a drastically reduced degrees of freedom. 
These models are made in mostly phenomenological ways 
to qualitatively have a few, selected aspects of 
turbulence. 
Celebrated models include the Burgers' equation \cite{burg} and 
the shell models \cite{Bi}, which are certainly more amenable 
to analytical and numerical studies. 
Another family tree of these models stems from the Constantin-Lax-Majda (CLM) model 
\cite{clm}. 
We find a turbulent solution of this model family for the first time
and discuss its relevance to two-dimensional Navier-Stokes turbulence. 

The CLM model was introduced to study the problem of putative finite-time blowup of 
the incompressible Euler equation in three dimensions, that is, whether or not the 
vorticity becomes infinite in a finite time starting from a smooth initial condition \cite{mb}.
The CLM model is a one-dimensional partial differential equation which 
models the three-dimensional (3D) inviscid vorticity equation. 
The velocity gradient in the vortex-stretching term is modeled 
as the Hilbert transform of the scalar vorticity to incorporate 
the Biot-Savart nonlocality. But the model omits the advection term.
Significantly, 
the model has an explicit analytic solution blowing up in a finite time.
The natural question was soon addressed: does the (hypo-) viscous effect, let us say 
adding the Laplacian (or various orders thereof) of the model vorticity, 
prevent a solution from blowing up? 
Unfortunately the answer was no \cite{scho, saka};
The vorticity
invariably
blows up in finite time however large order of the Laplacian is added.
This is in contrast to the existence of 3D Navier-Stokes (NS) solutions
with the Laplacian of order larger than $5/4$ \cite{rs}.

This indicates that the vortex stretching in the CLM equation is modeled somewhat 
excessively. Given that the viscosity is not enough to suppress the blowup,
what if the omitted advection term is retained in the CLM model? 
This was first considered by De Gregorio \cite{dg1, dg2} and it
is numerically shown that there exists a unique smooth solution to the CLM 
equation with the advection term globally in time \cite{oswn}, 
implying that the advection term is able to prevent the blowup,
although a rigorous proof is unavailable as yet. 
Subsequently, in order to study balance between the advection and
stretching terms for the existence of solutions,
the generalized Constantin-Lax-Majda-De Gregorio (gCLMG) model, 
$\partial_t \omega + a u \partial_x \omega =  \omega \partial_x u$,
was introduced in \cite{oswn}. 
Here $a \in \mathbb{R}$ is a parameter discussed below, $\omega(x,t)$
denotes the model vorticity and the model velocity $u(x,t)$
is expressed in terms of the vorticity 
as $u = -(-\partial_{xx})^{-1/2}\omega$ and $\partial_x u = H(\omega)$
which is the Hilbert transform of $\omega$.
The parameters $a = 0$ and $a = 1$ correspond to the CLM model and  
the CLM model with advection term, respectively. Note also that, for general $a$, the Galilean 
invariance is lost.
Mathematically, the short-time existence of solutions of the gCLMG equation 
is proven for all $a$ \cite{oswn},
while it is conjectured that there exists
some critical value, $a_c \sim 0.6$, above which a solution exists globally 
in time \cite{oswdcds}.

Here is a twist: negative $a$ can be considered \cite{oswn} 
at the expense of the analogy of the gCLMG model to the 3D
vorticity equation.
Instead, we gain the conservation law: for $a < 0$,
it is shown easily that $\int \vert \omega(x, t)\vert^{-a} dx$ is conserved \cite{oswn}.
For $a = -1$, the blowup of $\partial_x \omega$ is proven rigorously \cite{ccf},
while for $a=-2$ only numerical evidence for the blowup is available.

With these mathematical facts about the gCLMG model, 
we address the following natural question:  does a viscous gCLMG
model work as a physical model of turbulence?
It is thus reasonable to consider the following forced viscous
gCLMG model: 
\begin{eqnarray}
\partial_t \omega + a u \partial_x \omega = \omega \partial_x u + \nu \partial_{xx}\omega + f,
\label{gCLMG}
\end{eqnarray}
where $\nu$ is the model kinematic viscosity and $f(x,t)$ represents an external force.
For $a = 1$, albeit the similarity to the 3D vorticity equation, 
our numerical result shows that
the model's energy concentrates on the forcing scale and the inertial range (IR) 
is not developed at all. This indicates that
the model is inadequate for NS turbulence in the IR.
The reason for this inadequacy can be that the model admits no conserved quantity such as the energy.
On the other hand, for $a = -2$, the (model) enstrophy becomes the conservative quantity of the inviscid
gCLMG model. Such quadratic conservation law is an essential element of NS turbulence as stressed, e.g., in \cite{f}.
In this respect, the gCLMG model is akin to the 2D NS equation rather than the 3D vorticity equation.
Therefore, 
our aim in this paper is to investigate whether Eq.(\ref{gCLMG}) with $a = -2$ serves as a model of the
enstrophy-cascade turbulence of the two-dimensional (2D) incompressible NS equations. 
Among features of the enstrophy-cascade turbulence \cite{km, t,g,b}, 
we focus on similarities and differences of
the energy spectrum, in particular, the Kraichnan's logarithmic correction \cite{k71,lk72}
to the Kraichnan-Leith-Batchelor (KLB) energy spectrum, $k^{-3}$ \cite{k67, l68, b69}, and
of the structure function of the vorticity. We also try to describe the gCLMG turbulence
from a dynamical system point of view.

\section{Numerical simulation of the model}
We numerically simulate the gCLMG equation (\ref{gCLMG}) for $a = -2$ 
in a periodic interval of length $2\pi$ with a standard dealiased spectral method by assuming 
null vorticity Fourier mode of the zero wavenumber using the fourth order Runge-Kutta scheme
and the same filtering of the round-off noise as \cite{oswn}.
To achieve a statistically steady state, the large-scale forcing $f(x, t)$ is set to be random:
its Fourier coefficient $\hat{f}(k, t)$ is non zero
only for the forcing wavenumbers $k = \pm 1$, whose real and imaginary parts are set to 
Gaussian, delta-correlated-in-time and independent random variables with zero mean and 
variance, $\sigma_f^2$.
Specifically, with keeping $\sigma_f = 1.0\times10^{-2}$
(the average enstrophy-input rate is thus $2\sigma_f^2 = 2.0\times 10^{-4}$),
we vary the kinematic viscosity as
$\nu = \nu_0 \times 4^{-m} ~(m = 0, 1, 2)$ with $\nu_0 = 2.5\times10^{-5}$.
The corresponding time step, $\Delta t$, and the number of grid points, $N$, are
$(\Delta t, N) = (2.5\times10^{-4} \times 2^{-m}, 2^{13 + m})$.

\begin{figure} 
 \includegraphics[scale=0.65]{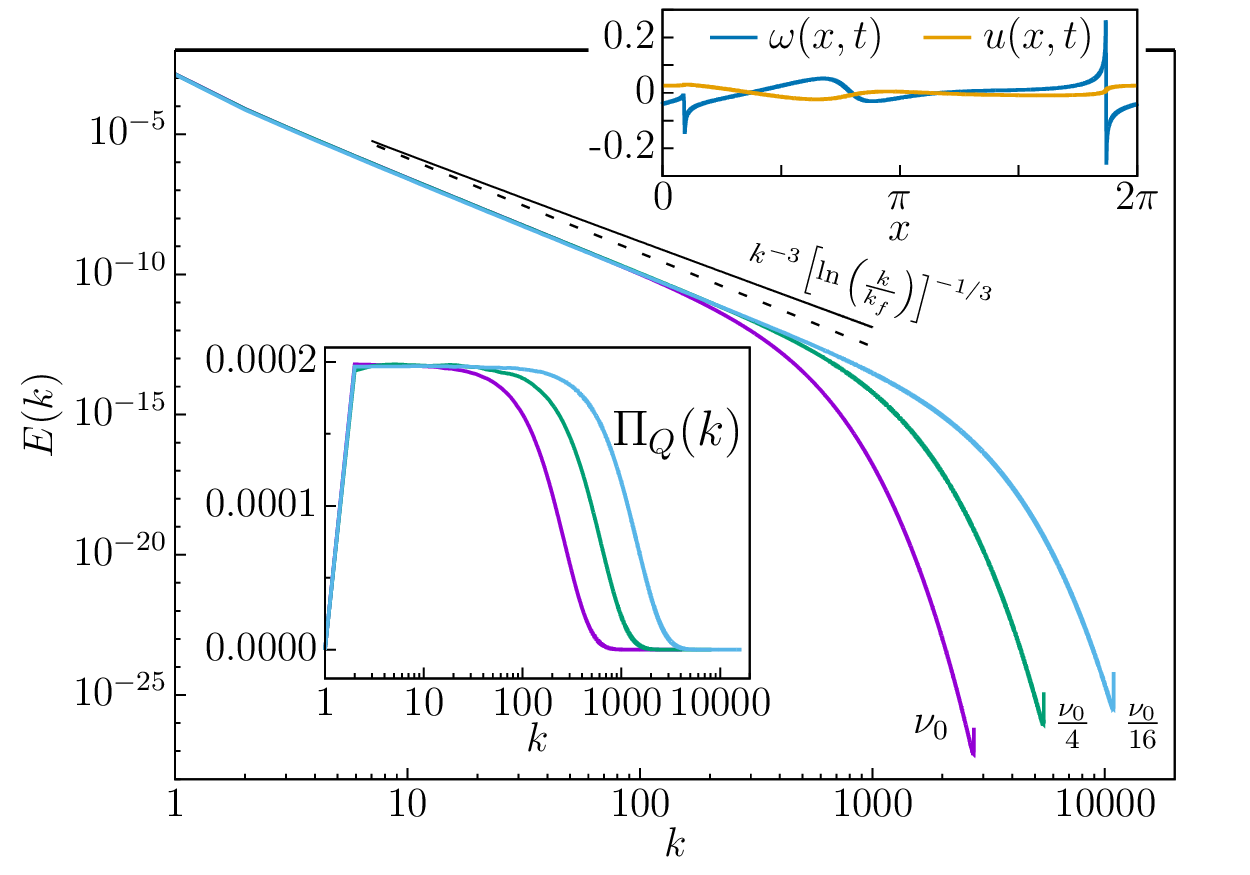}
 \caption{\label{spec}%
\textit{Top inset}: a typical snapshot of the vorticity with pulse structures
and the velocity of the randomly forced gCLMG equation.
\textit{Main figure}: the time-averaged energy spectra, $E(k)$, of
the forced gCLMG equation for three values of the viscosity.
The dashed line is Eq.(\ref{e}) for the inertial range.
\textit{Bottom inset}: the corresponding time-averaged enstrophy fluxes, $\Pi_Q(k)$.
The plateau value is equal to the average enstrophy input rate.}
\end{figure}
In the top inset of Fig.\ref{spec}, we show an example of the vorticity and velocity from 
the middle viscosity case. It is seen that 
the gCLMG turbulent state is characterized by several pulses of the vorticity, 
which move randomly and sometimes merge and emerge.
The energy spectra, $E(k) = \sum_{k \le |k'| < k + 1} |\hat{u}(k', t)|^2 / 2$ ($\hat{u}(k', t)$ is the velocity Fourier coefficient),
for the three cases are plotted in Fig.\ref{spec}. 
It exhibits the IR and dissipation range.
Indeed, as shown in the bottom inset of Fig.\ref{spec}, the enstrophy flux, $\Pi_Q(k, t)$, in the Fourier space  
of the gCLMG equation \cite{defflux} has a plateau which extends to larger wavenumbers
as we diminish the kinematic viscosity. 
The enstrophy flux is defined as 
$\Pi_Q(k, t) = \sum_{\ell (\ell \ge k)} \sum_{k' ~(|k'| = \ell)} \sum_{p, q ~(p + q = k')} {\rm Im}[\hat{\omega}^*(k', t) (aq - p)\hat{u}(p, t) \hat{u}(q, t)]$, where $\hat{\omega}(k', t)$ is the vorticity Fourier coefficient and $^*$ denotes complex conjugate.
Here we regard the plateau region as the IR of the gCLMG turbulence. 
Having obtained evidence of the enstrophy cascade in the forced gCLMG turbulence,
however,
we observe that the energy spectrum $E(k)$ in the IR
has a visible correction to the KLB law, $E(k) \propto \beta^{2/3} k^{-3}$,
where $\beta$ is the average enstrophy dissipation rate.
This law is the simplest dimensional-argument result on the enstrophy-cascade spectrum
in the IR \cite{k67, l68, b69}. 
Moreover, the gCLMG spectrum deviates also 
from the Kraichnan's prediction with the logarithmic correction 
about the 2D NS enstrophy-cascade turbulence,
$k^{-3} [\ln (k/k_f)]^{-1/3}$ \cite{k71, lk72},
where the forcing wavenumber is $k_f = 1$ here.
If we fit $E(k)$ with a pure power law, $k^{-\gamma}$, without the logarithmic correction, 
the exponent is close to $\gamma = 3.5$.
Nevertheless, we observe that dependence of $E(k)$ on $\beta$ is consistent
with the KLB law from collapse of $E(k) \beta^{-2/3}$ onto a single curve
in the inertial range for various enstrophy dissipation rates (figure not shown).
This indicates that the scaling behavior of the model is determined by the enstrophy cascade.
The scaling of the energy spectrum will be examined in terms of 
the vorticity structure function. 

\begin{figure} 
\includegraphics[scale=0.65]{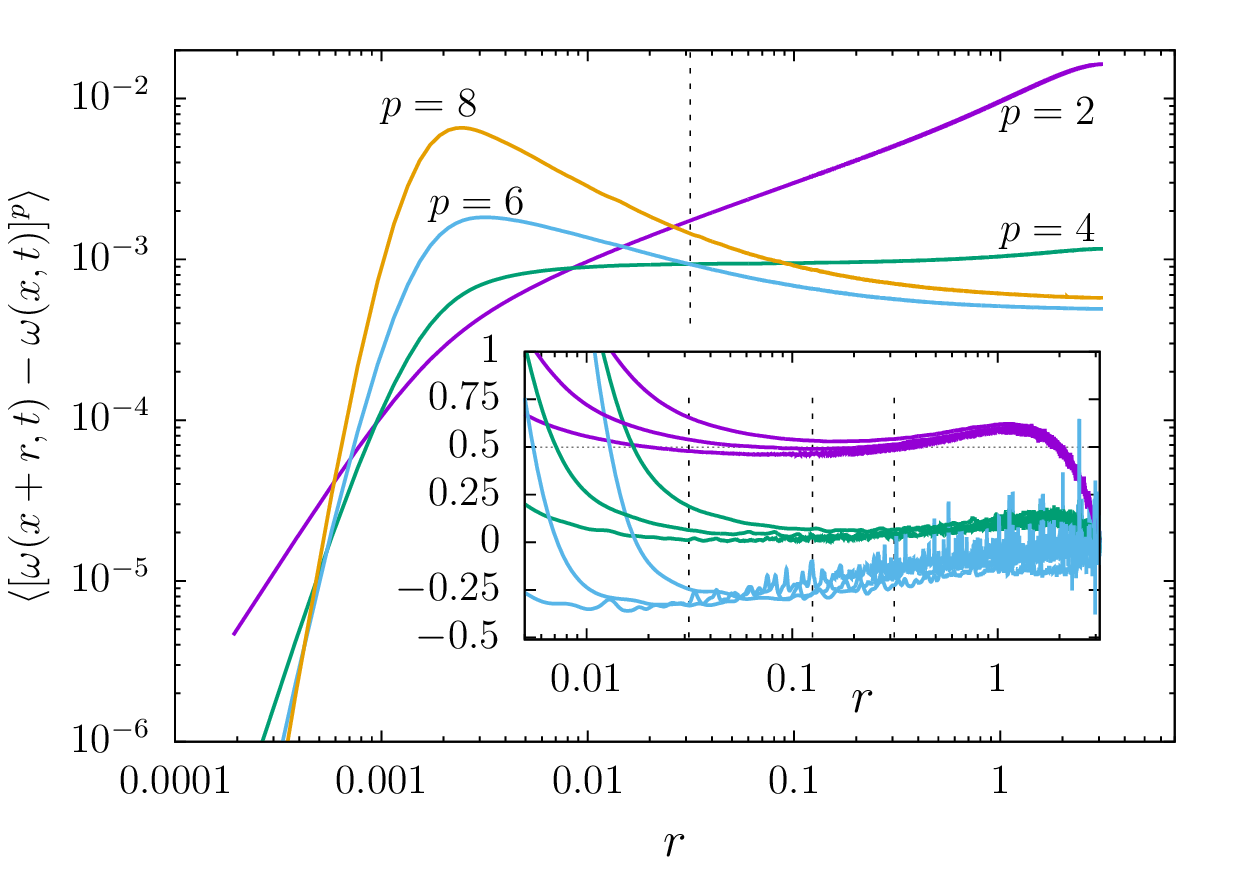}
\caption{\label{stfo} Even order ($p = 2, 4, 6, 8$) structure functions of the vorticity 
for the smallest viscosity case. 
The vertical dashed line corresponds to $r = 2\pi/200$. 
The inertial range is here estimated as $2\pi/200 \le r \le \pi$ from the bottom inset of Fig.\ref{spec} in which  
the plateau extends from $2 \le k \le 200$. 
\textit{Inset}: logarithmic local slopes of the structure functions ($p = 2, 4, 6$ from top to bottom) 
for the three viscosity cases. Oscillations tend to be reduced as we increase 
the number of statistical samples (the slope for $p=8$ is not shown because 
they are so large). 
The vertical dashed lines correspond to the respective 
smallest scales in the IR.} 
\end{figure} 
Now we move to the $p$-th order structure function of the vorticity, 
$S_p(r) = \langle (\delta_r \omega)^p \rangle$, where 
$\delta_r\omega = \omega(x + r, t) - \omega(x, t)$ and $\langle \cdot \rangle$ denotes
space-time average.
In Fig.\ref{stfo}, $S_p(r)$ are shown in the log-log coordinates.
It appears that $S_2(r)$ is close to $r^{0.5}$, which is consistent 
with $E(k) \propto k^{-3.5}$. 
However the logarithmic local slope, 
$[\ln S_2(r + \Delta r) - \ln S_2(r)] / [\ln (r + \Delta r) - \ln r]$
($\Delta r$ being the grid size),
does not converge to a unique constant value in the IR as we decrease
$\nu$.
One possibility is that the power-law exponent of $S_2(r)$ in the IR depends on
$\nu$ (similar dependence was found for the velocity structure function of 
the 2D enstrophy-cascade turbulence \cite{cg}). 
Another possibility is that the dominant part of $S_2(r)$ 
is not a pure power law but with a correction.
For the fourth order, it is seen that $S_4(r)$ is nearly constant in the IR.
Now let us assume that the even-order structure function is a pure power law
with exponent $\zeta_{2n}$ to the leading order, i.e., $S_{2n}(r) \propto r^{\zeta_{2n}}$.
Then the relation $\zeta_2 > \zeta_4$, as indicated in Fig.\ref{stfo}, 
implies that the vorticity cannot be bounded 
(see Sec.~8.4 of Ref.\cite{f}). 
With a given enstrophy input, $\omega$ of the gCLMG turbulence 
can be infinite for $\nu \to 0$ as we will see later.
This is in contrast to the velocity of the incompressible fluids. 
Conversely, if we consider that $\omega$ should be bounded,
then $S_{2n}(r)$ is not pure power law but with a correction.
We now have two scenarios: 
(i) the pure power law, $E(k) \propto k^{-3.5}$ and $S_2(r) \propto r^{0.5}$, leading to infinite vorticity
and 
(ii) the power law with a correction, $E(k) \propto k^{-3}\times$(non-power-law correction) 
and $S_2(r) \propto$(non-power law part), with bounded vorticity. 
We will later present evidence for (ii) while the infinite vorticity as $\nu \to 0$
is indicated simultaneously.

For higher orders, we observe 
that $S_6(r)$ and $S_8(r)$ have a peak in the dissipative range
and 
that they are decreasing functions in the IR.  
The peaks are due to the biggest pulse of the vorticity 
(the rightmost one in the top inset of Fig.\ref{spec}) as expected.
The scale of the peak, $r \simeq 0.002$, coincides with the typical
width of the biggest pulse. 
About the decreasing behavior in the IR, it is unlikely to be an artefact
caused by the peak since the decreasing region extends for smaller $\nu$ 
as seen in the bottom inset of Fig.\ref{stfo}. 
Then do $S_6(r)$ and $S_8(r)$ have negative scaling exponents in the IR,
as discussed for the 2D enstrophy-cascade turbulence \cite{ey}? 
Our answer is yes, but with non power-law corrections since 
the logarithmic slopes (here shown only for $S_6(r)$) are not 
satisfactorily flat. 
The negative exponents are consistent with the singular behavior of 
the vorticity (see \cite{f}). 
We conclude that $E(k)$ and $S_p(r)$ of the gCLMG turbulence
in the IR are not described by a pure power law but with certain corrections
and that $S_p(r)$'s behavior indicates a singularity of the vorticity.

Now we compare the above results of the gCLMG turbulence 
with those obtained for the 2D NS enstrophy-cascade turbulence 
(henceforth 2D turbulence).
We limit ourselves here to experimental and numerical results of 
the 2D turbulence having only the enstrophy-cascade IR
as a statistically steady state (not a decaying state).
Concerning the energy spectrum, the difference from the KLB $k^{-3}$ scaling
is more measurable in the gCLMG turbulence than in the 2D 
turbulence. The spectrum of the latter is occasionally fitted with the power law
$k^{-3.3}$ \cite{g,b,cee,cg,rae}. 
Regarding the vorticity structure functions of the 2D turbulence,
the result obtained in the laboratory experiment \cite{pct}
showed that those of even orders up to $10$ in the IR have power-law exponents 
indistinguishable from zero.
This is consistent with the theoretical results
beyond dimensional analysis of the 2D turbulence \cite{fl, ey}.
We note that in \cite{pct} 
no indication of $\zeta_2 > \zeta_4$ was observed, assuming
that the structure function in the IR is a power-law function.
Numerical results of the 2D turbulence showed that $S_2(r)$ 
is logarithmic without a power law and that $S_4(r)$ is a decreasing 
function \cite{vl}, however.
Switching to the velocity, we observe that both the even-order velocity structure functions 
and the even-order moments of the second-degree increments of the velocity \cite{fra}
of the gCLMG turbulence are not power-law functions in the IR as indicated by 
their logarithmic local slopes without plateau.
This is in contrast with the 2D turbulence \cite{bclvv, cg}.
Therefore, the gCLMG turbulence obeys distinct statistics    
from the 2D turbulence despite the analogous enstrophy cascade.
However both have statistics with non power-law type corrections in common.
Further details will be studied with a stationary solution of Eq.(\ref{gCLMG})
as follows.

\section{Stationary solution}
\begin{figure} 
\includegraphics[scale=0.65]{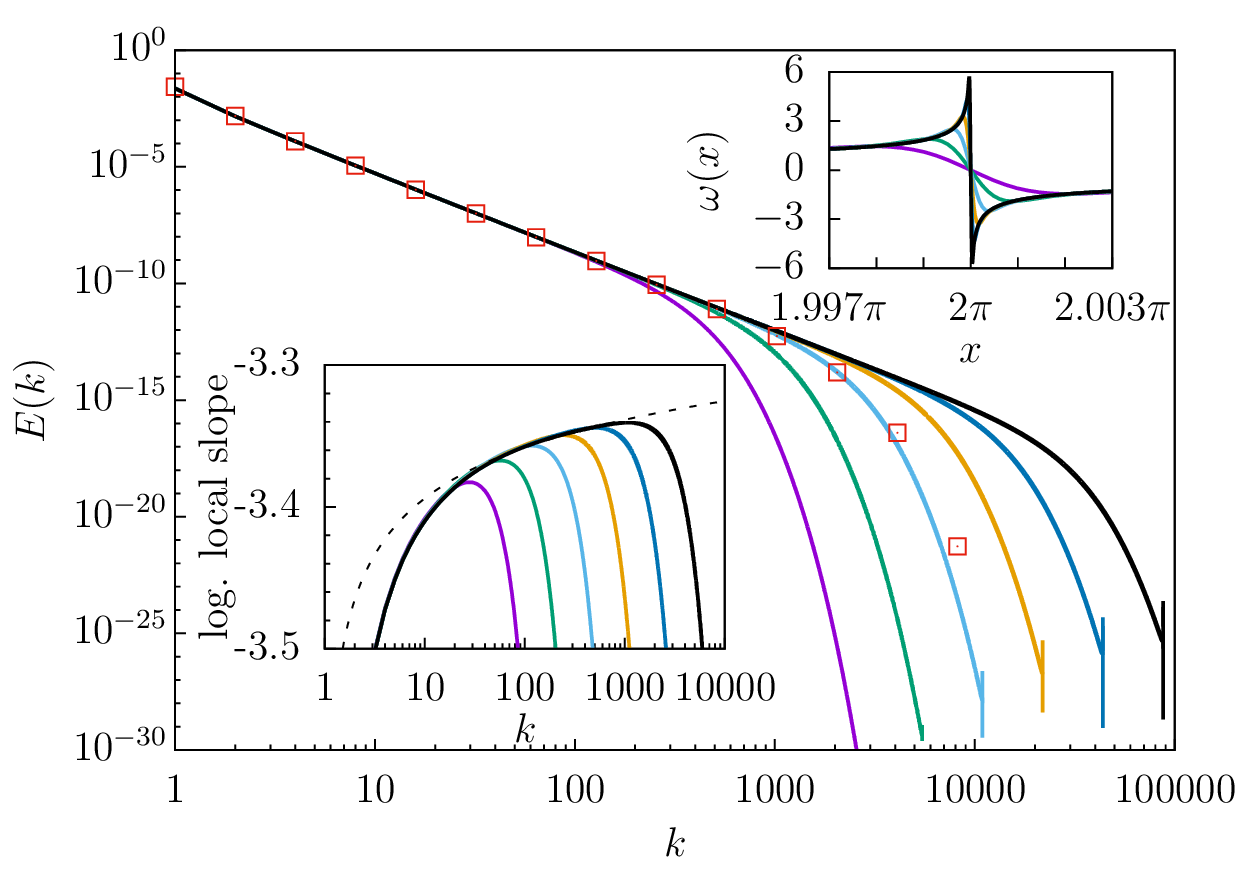}
 \caption{\label{statio} The energy spectra of the stationary solutions 
to the gCLMG equation with the deterministic forcing for various $\nu$'s. 
The squares represent data (multiplied by $20$) of the randomly forced case 
with $\nu_0 / 8$ shown in Fig.\ref{spec}.
\textit{Left inset}: the logarithmic local slopes of the energy spectra and
Eq.(\ref{a}) as the dashed line.
\textit{Right inset}: the vorticity pulses of the stationary solutions.
}
\end{figure}

Surprisingly, the forced gCLMG equation has a stable stationary solution whose
energy spectrum is indistinguishable to that of the gCLMG turbulence in the IR,
as shown in Fig.\ref{statio}.
We find it incidentally when we change the random forcing 
to a deterministic and stationary one, $f(x, t) = C_0 \sin x$, 
in order to study dependence of the statistics on the large-scale forcing.
Thus it demonstrates the striking forcing dependence.
We here fix the forcing amplitude $C_0 = -0.1$ and 
vary $\nu$ as $10^{-4} \times 4^{-m} ~(m = 0, 1, \ldots, 5)$. 
We use the same spectral method and time stepping as before 
with the corresponding time step $\Delta t = 2.5\times 10^{-4} \times 2^{-m}$ 
and the number of grid points $N = 2^{13 + m}$. 
The stationary solution has one vorticity pulse as depicted in the right inset of 
Fig.\ref{statio}, which is likely to converge to a singular function (infinite vorticity)
as $\nu \to 0$ (analogously nonsmooth vorticity has been observed in the 2D stationary 
Kolmogorov flow for small $\nu$ \cite{o96}).
Its peak value and width are numerically found to scale
with $\nu^{-0.2}$ and $\nu^{0.6}$, leading to the enstrophy dissipation rate
independent on $\nu$.

The identity of the energy spectrum of the stationary solution 
to that of the gCLMG turbulence in the IR implies 
that the functional form of the turbulent $E(k)$ can be studied with 
the stationary solution. 
In the left inset of Fig.\ref{statio}, we plot
the plateau-less logarithmic local slope of $E(k)$, demonstrating
no pure power law in the IR. 
We then make the following ansatz in the IR, 
\begin{equation}
 \frac{\ln \frac{E(k + \Delta k)}{E(k)}}
  {\ln \frac{k + \Delta k}{k}}
  \simeq 
k \frac{d}{dk} \ln E(k)
  = -c_0 - c_1 \left[\ln \mbox{$\left(\frac{k}{k_f}\right)$}\right]^\delta,
\label{a}
\end{equation}
although small discrepancy among $\nu$'s are present
(but not visible in the bottom inset of Fig.\ref{statio}).
For $\delta \ne -1$, integration of Eq.(\ref{a}) leads 
to the expression of the energy spectrum,
\begin{equation}
E(k) \propto k^{-c_0} \exp\left\{- c_2 \left[\ln \mbox{$\left(\frac{k}{k_f}\right)$} \right]^{\theta}\right\},
\label{e}
\end{equation}
where $c_2 = c_1 / (1 + \delta)$ and $\theta = 1 + \delta$.
Here we fix the first parameter as $c_0 = 3$ in view of the enstrophy cascade
and our previous observation of $E(k)$'s dependence on the enstrophy dissipation rate.
The rest of the parameters are estimated 
as $c_1 = 0.442$ and $\delta = -0.138$ by fitting Eq.(\ref{a}) 
in the range $20 \le k \le 1000$, resulting in $c_2 = 0.513$ and $\theta = 0.862$.
So far Eqs.(\ref{a}-\ref{e}) are empirical (but $\theta = -1$ case can 
be obtained with the incomplete self-similarity \cite{bc}).
The log-corrected spectrum of the form $E(k) \propto k^{-3} \ln^{\alpha}(k / k_f)$,
corresponding to $c_1 = -\alpha$ and $\delta = -1$,
does not yield a better fit however we adjust $\alpha$, implying
that this type of the correction is not suitable for 
the gCLMG turbulence.

\begin{figure} 
\centerline{
\includegraphics[scale=0.22]{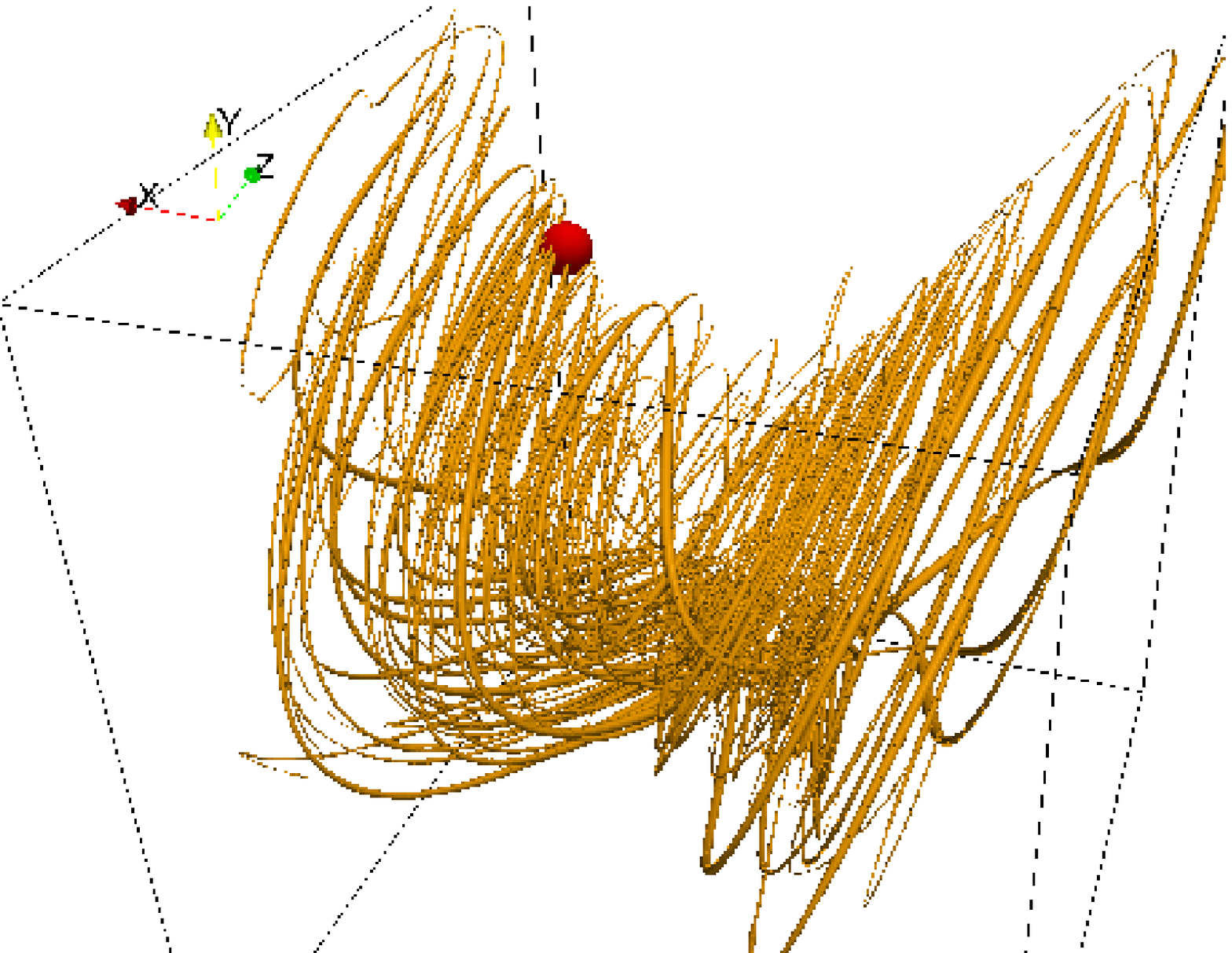} 
\hspace{0.1cm}
\includegraphics[scale=0.22]{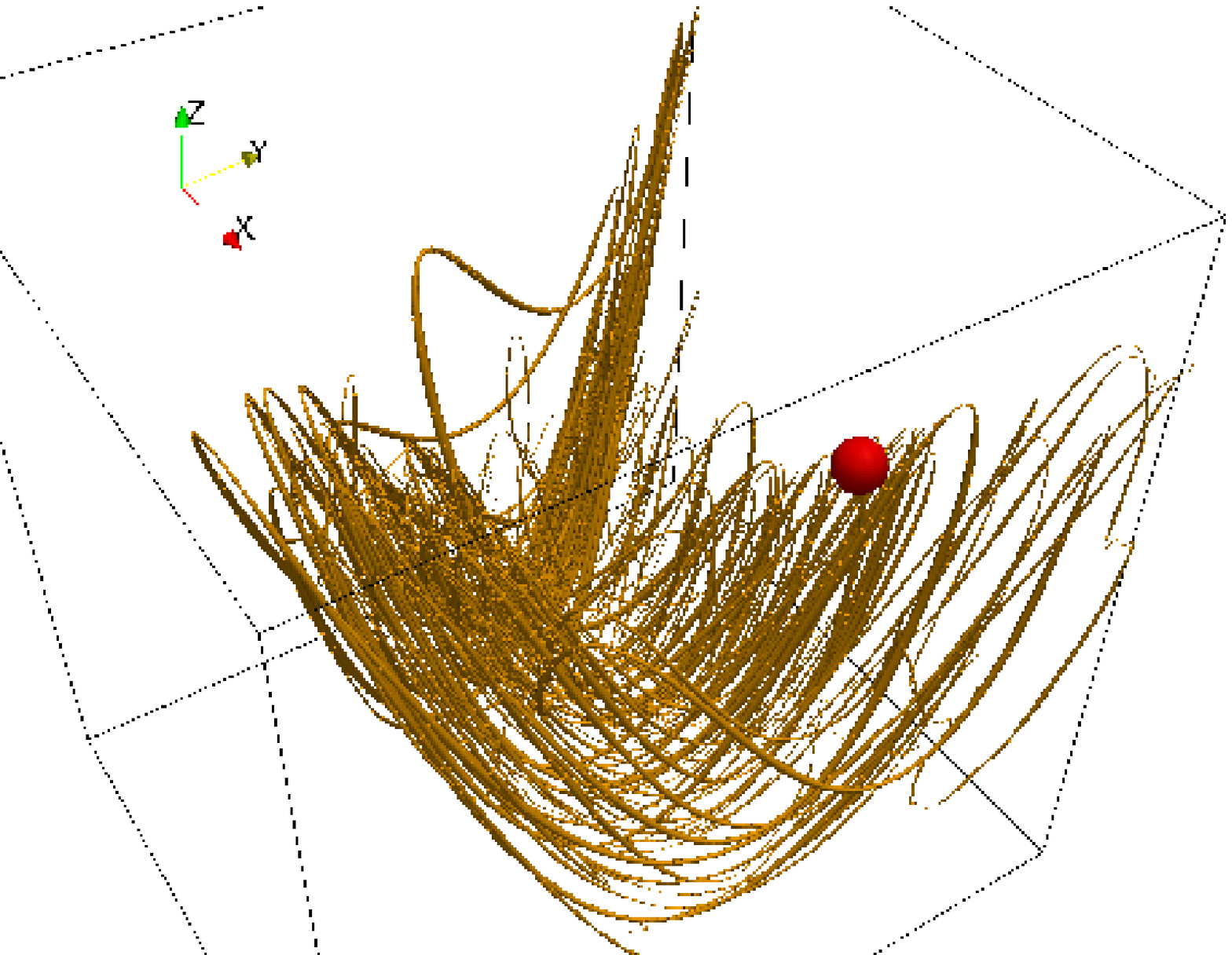}
}
\centerline{
\includegraphics[scale=0.22]{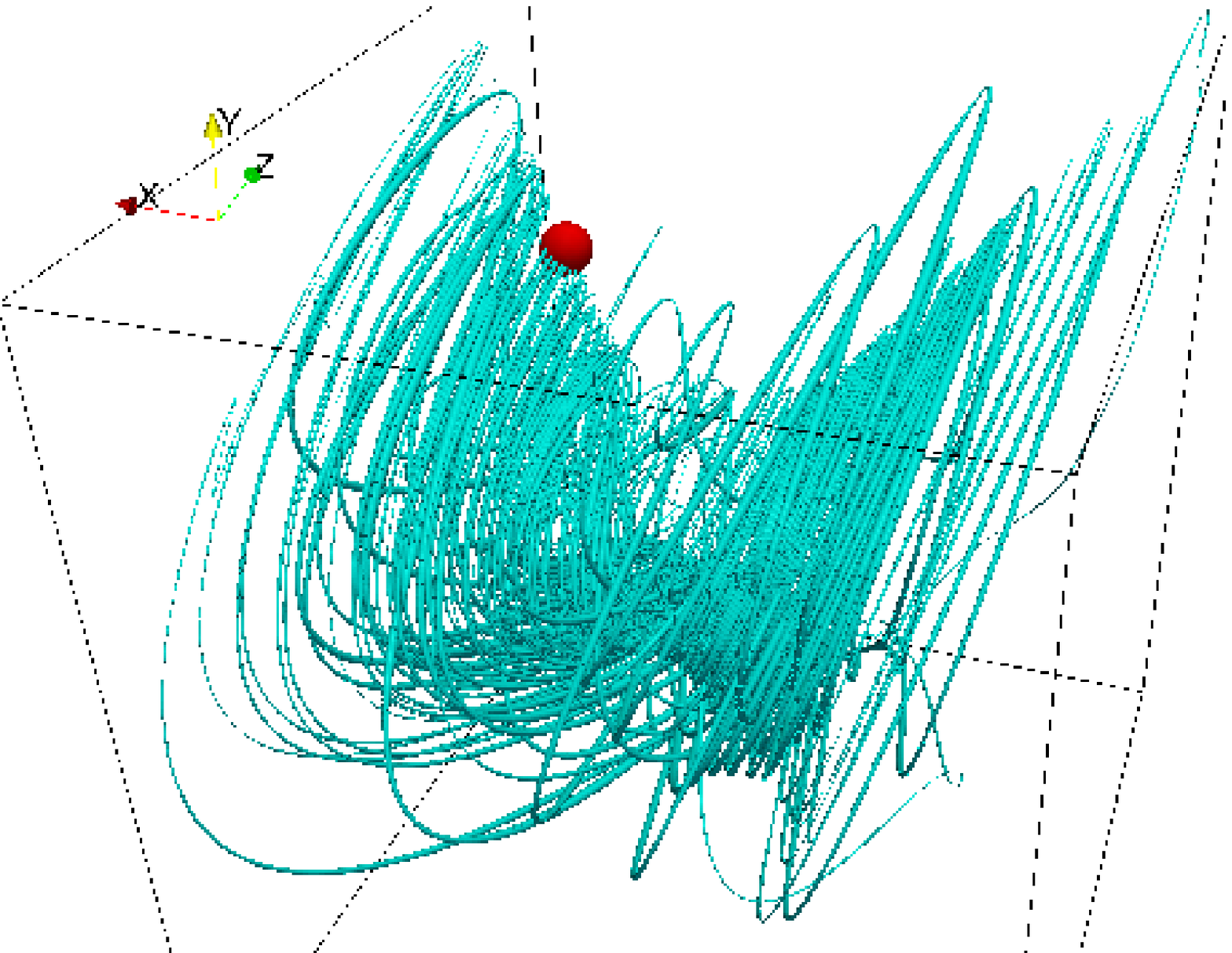}
\hspace{0.1cm}
\includegraphics[scale=0.22]{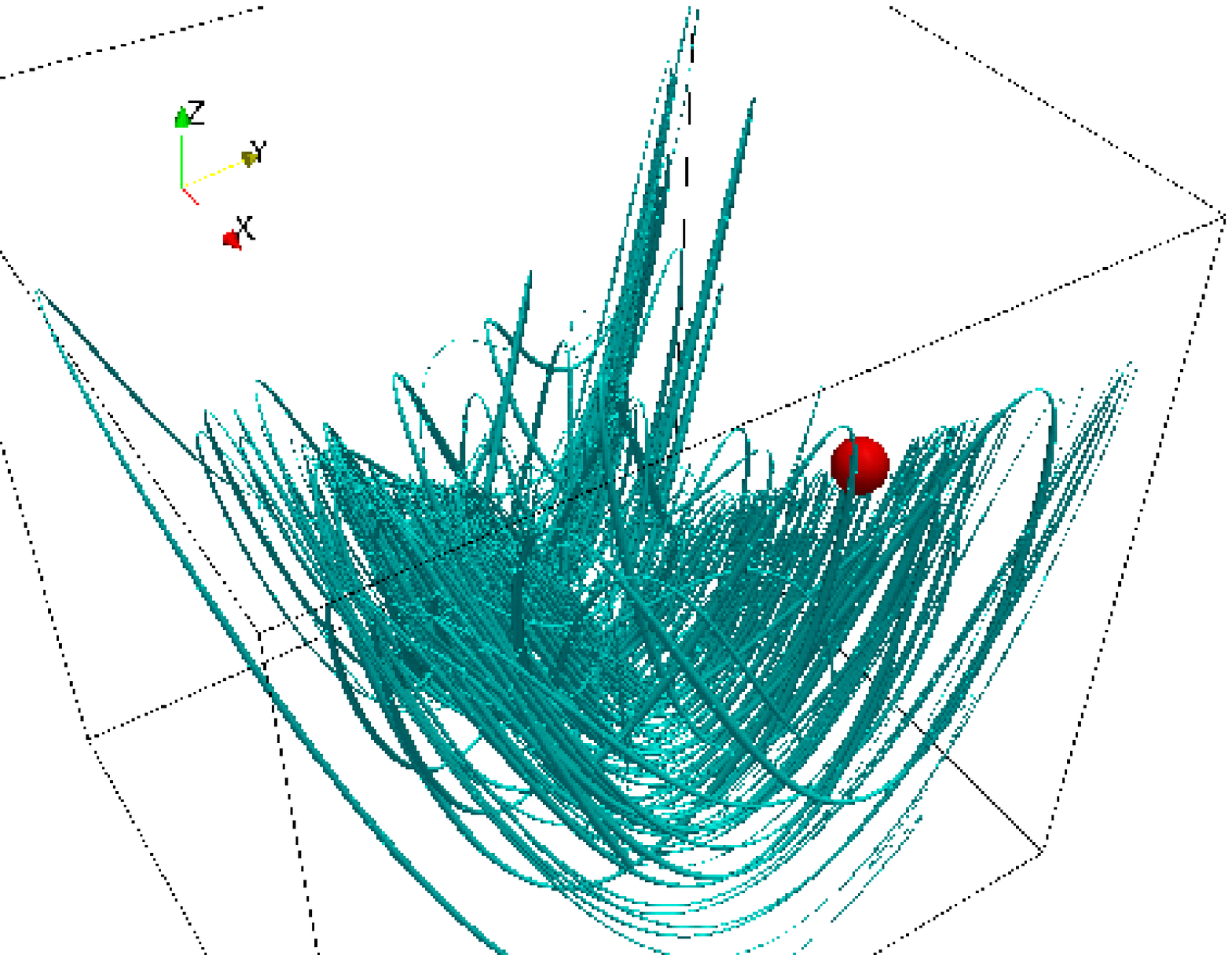}
}
\caption{\label{att} Self-similarity of the phase-space orbits of
${\rm Im}[\hat{\omega}(k, t)]$
of the randomly forced gCLMG turbulence ($\nu = \nu_0 / 4$ case)
for $k = 4, 8, 16$ (top) and $k = 32, 64, 128$ (bottom) as $(x, y, z)$ variables from 
two angles of view.
The point corresponds to the stationary solution with the same enstrophy and viscosity.}
\end{figure}
To see further relation of the stationary solution to 
the gCLMG turbulence from a dynamical system point of view,
we plot in Fig.\ref{att} an orbit of the randomly forced case
in the three dimensional 
subspace of the phase space, which is defined as
$(
{\rm Im}[\hat{\omega}(k_1, t)] / \Omega(k_1) - 1,\, 
{\rm Im}[\hat{\omega}(k_2, t)] / \Omega(k_2) - 1,\, 
{\rm Im}[\hat{\omega}(k_3, t)] / \Omega(k_3) - 1
)$. 
Here ${\rm Im}[\hat{\omega}(k, t)]$ 
and $\Omega(k)$ are the imaginary part of the vorticity Fourier coefficient 
of the randomly forced gCLMG equation and the stationary solution, respectively.
We take wavenumber triplet $(k_1, k_2, k_3)$ as $(4, 8, 16)$ and $(32, 64, 128)$ 
which are in the IR except for $128$. Hence we look at the orbit
scaled with the stationary solution from the two ranges of scales.
We observe that the orbit is within a thin surface 
with the stationary solution located on one edge of this attracting set.
Significantly, the orbits in the two scale ranges are similar. 
With this self-similarity of the orbit, modeling of the gCLMG turbulence 
with a few degrees of freedom is conceivable.

\section{Summary}
We numerically studied the gCLMG equation Eq.(\ref{gCLMG}) with $a = -2$ 
with a large-scale forcing.
The random forcing generates the turbulent state analogous
to the 2D NS enstrophy-cascade turbulence due to existence of 
the inviscid quadratic invariant, the enstrophy. 
The deterministic forcing yields the stable stationary solution 
which is spectrum-wise relevant to the gCLMG turbulence. 
The statistical laws of the gCLMG turbulence, such as the measurable corrections 
to the KLB law, are quantitatively different from the 2D turbulence.
However, qualitatively, statistics not characterized by simple power laws
are common.
Therefore we expect that the simpler gCLMG model provides insight
in theoretical study of these subtle statistics, which are not captured
by dimensional analysis.
Currently, we failed to substantiate our expectation although
the stationary solution should facilitate analytical study.
Another remarkable aspect of the gCLMG solutions is the strong indication of
infinite vorticity with a finite enstrophy dissipation rate as $\nu \to 0$, 
possibly an inheritance from the CLM model.
This implies that the gCLMG model is an interesting testing ground for  
investigating relation between singularity and turbulence statistics.
However, we note that the finite-limit of the enstrophy dissipation rate is not shared
with the 2D enstrophy-cascade turbulence (see, e.g.,\cite{d1, d2}).
For different negative $a$'s in Eq.(\ref{gCLMG}), we found that the similar results
to the $a=-2$ case holds, which will be reported elsewhere. 

\section*{Acknowledgments}
We acknowledge delightful discussions
with Koji Ohkitani, Hisashi Okamoto and Shin-ichi Sasa
and the support by Grants-in-Aid for Scientific Research 
KAKENHI (B) No.~26287023 from JSPS.

\end{document}